\documentclass{article}

    \PassOptionsToPackage{numbers, compress}{natbib}

    \usepackage[final]{neurips_2024}

\usepackage[utf8]{inputenc} 
\usepackage[T1]{fontenc}    
\usepackage{url}
\usepackage{ulem}

\usepackage{breakurl}
\usepackage[breaklinks]{hyperref}
\usepackage{booktabs}       
\usepackage{amsfonts}       
\usepackage{nicefrac}       
\usepackage{microtype}      
\usepackage{xcolor}         
\usepackage{pifont} 

\newcommand{\cmark}{\ding{51}}
\newcommand{\xmark}{\ding{55}}

\newcommand{\etal}{\textit{et al}.\ }
\usepackage{algorithm}
\usepackage{algorithmic}

\usepackage{amsmath}
\usepackage{amssymb}
\usepackage{mathtools}
\usepackage{amsthm}
\usepackage{stix}
\usepackage[capitalize,noabbrev]{cleveref}
\usepackage{multirow}
\usepackage{colortbl}
\usepackage{subfigure}
\usepackage{xspace}
\usepackage{caption}
\usepackage{enumitem}
\usepackage{wrapfig}
\theoremstyle{plain}

\theoremstyle{definition}

\theoremstyle{remark}

\newcommand{\ouralg}{\texttt{Bileve}\xspace}

\newcommand{\KeyGen}{\mathsf{KeyGen}}
\newcommand{\Embed}{\mathsf{Embed}}
\newcommand{\Verify}{\mathsf{Verify}}
\newcommand{\true}{\mathtt{true}}
\newcommand{\false}{\mathtt{false}}
\newcommand{\sk}{\mathsf{sk}}
\newcommand{\pk}{\mathsf{pk}}

\newcommand{\Sign}{\mathsf{Sign}}
\newcommand{\SLS}{\texttt{SLS}\xspace}

\usepackage{xcolor}  
\definecolor{darkred}{rgb}{0.7, 0, 0}  
\definecolor{darkgreen}{rgb}{0, 0.7, 0}

\usepackage{tikz}
\newcommand{\circlen}[1]{
  \tikz[baseline=-0.75ex]\node[circle, minimum size=0.3cm, inner sep=1pt, text=darkred, draw=darkred] {#1};
}

\newcommand{\circlednum}[1]{%
  \tikz[baseline=-0.75ex] \node[circle, fill=darkred, text=white, minimum size=0.5em, inner sep=1pt] {#1};%
}
            
\definecolor{darkblue}{rgb}{0, 0, 0.5}
\hypersetup{colorlinks,linkcolor={darkblue},citecolor={darkblue},urlcolor={darkblue}}

\usepackage[textsize=tiny]{todonotes}

\title{\ouralg: Securing Text Provenance in Large Language Models Against Spoofing with Bi-level Signature }

\author{%
  Tong Zhou \\
  Northeastern University, Boston \\
  \texttt{zhou.tong1@northeastern.edu} \\
  \And
  Xuandong Zhao \\
  UC Berkeley \\
  \texttt{xuandongzhao@berkeley.edu} \\
  \AND
  Xiaolin Xu \\
  Northeastern University, Boston \\
  \texttt{x.xu@northeastern.edu} \\
  \And
  Shaolei Ren \\
  UC Riverside \\
  \texttt{shaolei@ucr.edu} \\
}

\begin{document}

\maketitle

\begin{abstract}
Text watermarks for large language models (LLMs) have been commonly used to identify the origins of machine-generated content, which is promising for assessing liability when combating deepfake or harmful content. While existing watermarking techniques typically prioritize robustness against removal attacks, unfortunately, they are vulnerable to spoofing attacks: malicious actors can subtly alter the meanings of LLM-generated responses or even
forge harmful content, potentially leading to the wrongful attribution of blame to the LLM developer. To overcome this, we introduce a bi-level signature scheme, \ouralg , which embeds fine-grained signature bits for integrity checks (mitigating spoofing attacks) as well as a coarse-grained signal to trace text sources when the signature is invalid (enhancing detectability) via a novel rank-based sampling strategy. Compared to conventional watermark detectors that only output binary results, \ouralg can differentiate 5 scenarios during detection, reliably tracing text provenance and regulating LLMs. The experiments conducted on OPT-1.3B and LLaMA-7B demonstrate the effectiveness of \ouralg in defeating spoofing attacks with enhanced detectability. Code is available at \url{https://github.com/Tongzhou0101/Bileve-official}.

\end{abstract}

\section{Introduction}
Watermarks have been envisioned as a promising method to differentiate content generated by large language models (LLMs) from human \cite{deepmind, kirchenbauer2023watermark, zhao2024provable, kuditipudi2023robust, aaronson, christ2023undetectable}.
It involves injecting statistical signals into the token sampling process utilizing a secret watermark key. Subsequently, the one who knows the key can verify the content's origin by assessing the presence of the predefined signal through a statistical test. 
Current watermarking schemes primarily focus on user-side concerns, striving to achieve robustness against watermark removal attacks (i.e., perturb the generated text to remove the watermark), thereby combatting academic dishonesty and other deceptive practices.

However, a critical vulnerability remains unaddressed in watermark design: \textit{spoofing attacks directed towards model owners}.  In these attacks, malicious actors attempt to falsely attribute content generated by humans or other models to the targeted model, with the aim of evading accountability\footnote{Tesla lawyers claimed that Elon Musk’s past statements about self-driving safety could be deepfakes, which is found suspicious by the court \cite{tesla}.} or damaging the model's reputation. A few recent works have identified two kinds of spoofing attacks targeting the LLM watermark by exploiting either its symmetric characteristic \cite{fairoze2023publicly, liu2024a} or learnability \cite{gu2023learnability,jovanovic2024watermark}, as detailed in Sec.~\ref{sec:spoof}.
Furthermore, we propose a new spoofing attack, named semantic manipulation, which enables attackers to alter the sentiment of generated content with minimal token modifications, as described in Sec.~\ref{sec:our_attack}. It assumes the most constrained
capabilities of attackers, where they only have access to the victim model’s detector. Due to the robustness of LLM watermarks against perturbations,  this attack can manipulate the originally helpful content into something harmful or offensive without compromising the detectability of the watermark, thus successfully achieving spoofing attacks.

Given the serious consequences of spoofing attacks, it is highly in demand to answer the question:
\textit{\textbf{How to avoid an LLM being wrongly blamed? }}
To solve this problem, we aim to design a watermark for LLM, which focuses more on the model owners' side instead of only watermarking on the users' side. To reliably identify the provenance of machine-generated content while being able to defend against the above spoofing attacks, a signature should have the following properties:
\begin{itemize}[leftmargin=*,itemsep=0pt, topsep=0pt]
    \item \textbf{Robust:} The signature remains capable of tracing the source of machine-generated text even when subjected to certain perturbations, ensuring it is not overly fragile or easily rendered ineffective.
    \item 
    \textbf{Unforgeable:} The signature is inherently resistant to being learned given the components utilized in its detection.     
    \item \textbf{Tamper-evident:} It should be able to check the integrity of the generated content, showing reliable tampering evidence to safeguard the interests of model owners. 
    \item \textbf{Transparent:} It is detectable without needing access to generation secrets or relying on a black-box API, allowing independent, reliable verification.
\end{itemize}

\begin{table}[t]
\centering
\caption{Comparison of different watermarking methods based on four desired properties. }
\label{tab:comparison}
\resizebox{0.83\linewidth}{!}{
\begin{tabular}{l|ccccc}
\hline
\hline
\textbf{Methods} & \textbf{Robust}  & \textbf{Unforgeable} & \textbf{Tamper-evident}&  \textbf{Transparent}  \\
\hline
Kirchenbauer \etal\cite{kirchenbauer2023watermark} & \cmark  & \xmark & \xmark &  \xmark    \\
Zhao \etal\cite{zhao2024provable} & \cmark  & \xmark & \xmark&  \xmark    \\
Kuditipudi \etal\cite{kuditipudi2023robust} & \cmark  & \xmark & \xmark & \xmark    \\
Liu \etal\cite{liu2024a} & \cmark  & \xmark  & \xmark  & \xmark  \\
Fairoze \etal\cite{fairoze2023publicly} & \xmark   & \cmark & \cmark &  \cmark   \\
\hline
\hline
\end{tabular}}
\vspace{-0.8cm}
\end{table}

Despite the critical importance,  achieving all desired properties in a single LLM watermark remains challenging, as even state-of-the-art (SOTA) designs cannot meet them all (see Tab.~\ref{tab:comparison}). Indeed, 
designing such a watermarking scheme involves a fundamental trade-off between defending against removal attacks and spoofing attacks. Specifically, being robust to removal attacks requires that the watermark's detectability remains unaffected by certain perturbations, while anti-spoofing demands sensitivity to perturbations to verify text integrity, distinguishing harmful content from genuine model output and tampered content.

To overcome the above challenges, we propose \ouralg, a novel sampling strategy by embedding a bi-level signature into generated tokens. At the coarse-grained level, we utilize statistical signals across the entire text to detect the presence of the watermark, ensuring robustness against perturbations. Concurrently, at the fine-grained level, we integrate content-dependent signature bits into each token to uphold content integrity, which leverages a digital signature scheme to ensure unforgeability, as the secret key required for watermark embedding will be securely held by model owners. This scheme enables transparent detection by allowing verification with a public key instead of embedding secrets, so independent parties can authenticate without proprietary details or a black-box API, ensuring reliable detection. And the tampering evidence will show when these two level detection results are not consistent.

Our contributions are threefold: 1) We uncover an advanced spoofing attack that exploits the robustness of SOTA watermarking schemes; 2) We introduce \ouralg, the first watermarking scheme to simultaneously ensure robustness and unforgeability by embedding a bi-level signature through a novel rank-based sampling strategy; 3) \ouralg is capable of distinguishing five distinct scenarios during the detection phase, effectively defeating spoofing attacks and serving as a promising tool to regulate LLM safety mechanism.

\section{Background and Related Works}
\subsection{Language Model Basics}
Let $\mathcal{M}$ denote a language model with a vocabulary $\mathcal{V}$ containing $K$ $:=|\mathcal{V}|$ tokens. To generate the next token $w_t$,  $\mathcal{M}$ will take prior tokens $w_{1:t-1}$ as the input and output a vector of logits ${l}^{(t)}$, which is transformed into a probability distribution $\mathcal{D}^{(t)}=(p^{(t)}_1,...,p^{(t)}_K)$ via the softmax function.
Then the sampling strategy is applied to determine how the model selects $w_t$ based on $\mathcal{D}^{(t)}$. One common sampling strategy is multinomial sampling, where $\mathcal{M}$ randomly selects the next token from $\mathcal{V}$ according to the probabilities $p^{(t)}_k$ assigned to each token. 
This process is repeated iteratively to generate a sequence of tokens.

\vspace{-0.3cm}
\subsection{LLM Watermarks}
Watermarks for model-generated texts are used to identify the provenance of the text, ensuring accountability in cases where generated content needs to be traced back to a specific LLM. 
The existing watermark schemes rely on the specialized decoding algorithm to embed statistical signals into generated contents, then enabling watermark detection via computing p-value~\cite{aaronson, kirchenbauer2023watermark, zhao2024provable, christ2023undetectable, kuditipudi2023robust}.
For instance, for generating the next token, one approach dynamically partitions the vocabulary into green and red lists based on its previous few tokens and a watermark key~\cite{kirchenbauer2023watermark}, 
then increasing the logits of green tokens to enhance their chance of being selected. During detection, the watermark detection key is used to count the number of green tokens in the text, with the calculated z-statistic indicating the existence of the watermark. Moreover, Zhao \etal \cite{zhao2024provable} simplify the scheme proposed in~\cite{kirchenbauer2023watermark} by fixing the green-red list for each token, demonstrating that their watermark is twice as robust to edit as~\cite{kirchenbauer2023watermark}.
Furthermore, unlike modifying logits, a distortion-free watermark is proposed to preserve the original text distribution~\cite{kuditipudi2023robust}. It leverages robust sequence alignment to align watermarked text to a watermark key sequence in the sampling phase, e.g., using exponential minimum sampling. 

However, these watermarking schemes only enable detection by individuals possessing the key, which doesn't facilitate transparent regulation. On one hand, making the key public is susceptible to attacks~\cite{liu2024a}. On the other hand, maintaining detection privately (e.g., via APIs) compromises reliability, as it functions as a black box, allowing the model owner to manipulate detection results.

\subsection{Spoofing Attacks}
\label{sec:spoof}
\begin{table}[t]
    \centering
    \caption{Three categories of spoofing attacks (ranked by attackers' capabilities from high to low).}
    \resizebox{\linewidth}{!}{%
    \begin{tabular}{lcl}
    \toprule
        \textbf{Methods}  & \textbf{Exploited Vulnerabilities} & \textbf{Attackers' Capabilities}\\ \midrule
        \cite{fairoze2023publicly,liu2024a}   & Symmetry & Know the secret key for embedding watermarks \\
        \cite{gu2023learnability,jovanovic2024watermark}   & Learnability & Get access to the victim model and query it multiple times\\
        Our attack   & Robustness & Only get access to the victim models' detector\\ \bottomrule
    \end{tabular}    }
    \label{tab:spoofing}
    \vspace{-0.3cm}
\end{table}

Spoofing attacks can fall into three categories based on the capability of attackers, and each of them exploits different vulnerabilities in SOTA watermarks, as summarized in Tab.~\ref{tab:spoofing}. First, due to the watermark embedding and detection process sharing the same secret key (i.e., symmetric schemes), the semi-honest detector knowing the secret key can embed the watermark to any content. Such a vulnerability has been discussed in \cite{fairoze2023publicly, liu2024a}, where they design asymmetric watermarking schemes so that detection does not rely on the key used for embedding. 
Specifically, \cite{fairoze2023publicly} proposes embedding watermarks using digital signature algorithms, ensuring that only model owners possess the secret key for watermark embedding while providing detectors with access to the public key for detection. However, it is easily broken once the message tokens are perturbed. Besides, it suggests using error-correcting encoding to improve robustness, which unfortunately would increase the risk of spoofing attacks and should not be adopted. 
Moreover, \cite{liu2024a} employs two distinct neural networks for watermark embedding and detection, leveraging an asymmetric scheme for public detection.

However, \cite{liu2024a} is based on \cite{kirchenbauer2023watermark} and thus can be learned as studied in \cite{gu2023learnability}, which is the second kind of spoofing attacks.  In particular, \cite{gu2023learnability} demonstrated that, by querying the victim model and collecting its watermarked samples, attackers can fine-tune an adversary model utilizing a sampling-based watermark distillation technique to learn the watermark (detailed in Appendix~\ref{sec:forgery}). The fine-tuned adversary model can respond to any malicious requests, with the response containing the watermark of the victim model. 
Recently, an independent work also proposed attacks by exploiting watermark robustness \cite{pang2024attacking}. Beyond techniques such as randomly inserting toxic tokens or modifying tokens to alter sentence accuracy, our attack leverages a reward model to guide targeted semantic manipulation. More importantly, while they suggest compromising watermark robustness to mitigate spoofing attacks, our work demonstrates how to achieve effective mitigation without sacrificing robustness.

\section{Potential Attack: Semantic Manipulation}

\subsection{Threat Model}

\textbf{Attackers' Objective.} Given text generated by the victim LLM, attackers seek to alter the semantic meaning of the text with minimal changes, transitioning it from something helpful or neutral to harmful or offensive.  Owing to the robustness of the existing watermarks, the watermark detector can still identify the presence of the watermark in the modified content. Consequently, the altered content erroneously attributes its origin to the victim model, potentially damaging the model's reputation.

\textbf{Attackers' Capability.} Contrary to existing spoofing attacks, our approach assumes the strictest attacker capabilities, where the adversary only gains access to the watermark detector, as outlined in Tab.~\ref{tab:spoofing}. These attackers lack knowledge of the secret key and are not required to query the victim LLM multiple times to acquire watermarked samples for training other adversary models. They may utilize public language models to enhance the efficiency of their attacks.

\subsection{Attack Method}
\label{sec:our_attack}

We identify the dilemma of being robust and unforgeable. In particular,  being robust indicates that the watermark should be preserved after perturbation, thus we can exploit this characteristic to design a novel spoofing attack.
Specifically, attackers can query the victim model with harmless prompts, and then use basic word replacement techniques to change its semantic meaning to be toxic or harmful. \textit{Due to the robustness properties of LLM watermarks, the detectability will not be compromised if the portion of word replacement is low.} 
Consequently, a detector cannot discern whether the content originated from the victim model or was manipulated by malicious actors. This highlights the limitation of current watermarks for auditing LLMs.

By exploiting the above observation, we propose a novel spoofing attack. Let $w^{\text{orig}}$ denote the original response of victim models, and $w^{\text{att}}$ represent its manipulated version. The goal is to generate $w^{\text{att}}$ that maximizes the change in sentiment while minimizing the Levenshtein distance between the original and manipulated responses. The problem can be formulated as follows:
\begin{equation}
\begin{aligned}
\max_{w^{\text{att}}} \Delta R = R(w^{\text{orig}}) - R(w^{\text{att}}),\;\;
\operatorname{s.t.} ~~ \text{LD}(w^{\text{orig}}, w^{\text{att}}) \leq \epsilon T
\end{aligned}
\label{eq:attack}
\end{equation}
Here, $\Delta R$ represents the sentiment change, defined as the difference between the reward scores (denoted by $R(\cdot)$) of the original and manipulated responses obtained by the reward model.\footnote{For example, we use the following reward model in our experiment: \url{https://huggingface.co/OpenAssistant/reward-model-deberta-v3-large-v2}} A lower score of the reward model indicates less alignment with human feedback, such as a toxic response. The Levenshtein distance, denoted by $\text{LD}(t_1, t_2)$, measures the minimum number of word 
edits required to transform text $t_1$ into text $t_2$. $T$ is the length of $w^{\text{orig}}$ and $\epsilon$ is the word edit budget. A trade-off exists in choosing $\epsilon$, wherein a larger value affords greater flexibility in manipulating the semantic meaning of $w^{\text{orig}}$, while a smaller value better preserves the detectability of the watermarks. To strike a balance, we opt for a larger $\epsilon$ to maximize semantic alteration and introduce a tuning factor $\alpha \in (0,1)$ to adjust $\epsilon$ in case the detectability is broken. 
Furthermore, instead of manually replacing the words in $w^{\text{orig}}$, attackers can simply leverage a powerful and accessible LLM (denoted as $Q$) to execute such attacks efficiently.  To enhance the generation quality while meeting the constraint, attackers can apply in-context learning by providing a few task demonstrations.
We summarize the algorithm with more details of task demonstrations in Appendix~\ref{app:attack}.


\section{Proposed Defense}
\begin{figure}[t]
    \centering
    \includegraphics[width=0.98\textwidth]{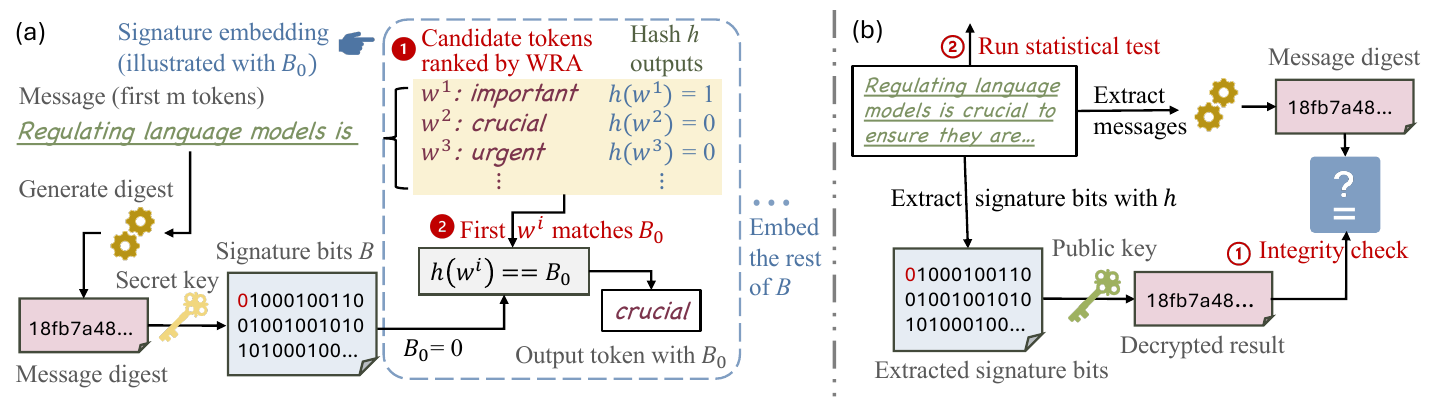}
    \caption{\textbf{Overview of \ouralg.} \textbf{(a) Embedding:} The first $m$ tokens from $\mathcal{M}$ form the message, which is signed using a secret key. Candidate tokens are selected via a rank-based strategy employing a Weighted Rank Addition (WRA) score, with a coarse-grained signal embedded. It then embeds the fine-grained signature by choosing the first candidate matching the designated signature bit. 
\textbf{(b) Detection:} We first extract the message-signature pair to conduct an integrity check using the public key. A statistical test is performed if necessary. }
    \label{fig:overview}
\end{figure}
With the above attack, in conjunction with other existing spoofing attacks, we can recognize the vulnerability of current watermark schemes. This underscores the importance of designing secure schemes to defend against spoofing attacks and achieve all properties listed in Tab.~\ref{tab:comparison}, ensuring reliable identification of text provenance.

\subsection{Single-level Signature}
To enable secure and reliable text attribution, we first examine the vulnerabilities exploited by attackers in conducting spoofing attacks, including symmetry, learnability, and robustness, as summarized in Tab.~\ref{tab:spoofing}. 
In particular, previous methods embed the statistical signal into the generated texts \cite{kirchenbauer2023watermark,zhao2024provable,kuditipudi2023robust, aaronson}, so as to identify the existence of a watermark during detection. Such statistical signal is consistent for every text, thus enabling the adversary model to learn the watermark rule can forge it. 
Therefore, to defend against spoofing attacks, we seek the opposite characteristics, i.e., \textit{asymmetric}, \textit{unlearnable}, and \textit{perturbation-sensitive}, in the solution.

We envision the digital signature as a promising solution as in~\cite{fairoze2023publicly}, where the scheme is defined as $\SLS=(\KeyGen, \Sign, \Embed, \Verify)$, where:
\begin{itemize}[leftmargin=*,itemsep=0pt, topsep=0pt]
    \item $\KeyGen \to (\sk, \pk)$ outputs a public key pair $(\sk, \pk)$, where $\pk$ is public while $\sk$ is held by the owner of model $\mathcal{M}$.
    \item $\Sign_\sk(\mathbf{m}) \to \boldsymbol{\sigma}$ uses $\sk$ to sign the digest of the message $\mathbf{m}$ := $w_{1:m}$ via the hash function $H$ (e.g., MD5) and obtain the signature $\boldsymbol{\sigma}=\sk \times H(\mathbf{m})$.
    \item $\Embed(\boldsymbol{\sigma}) \to w$ embeds signature into subsequent tokens following $w_m$, and outputs $w$ incorporating the message-signature pair ($\mathbf{m}, \boldsymbol{\sigma}$). 
    \item $\Verify_{\pk}(w) \to \left\{ \true,\false \right\}$ extracts $\boldsymbol{\sigma}=\sk \times H(\mathbf{m})$ from $w$ and verifies it using the public key $\pk$. If verification succeeds, it outputs $\true$; otherwise, it outputs $\false$. 
\end{itemize}

Unlike the digital signature methods, which typically attach signatures as metadata~\cite{subramanya2006digital,sagar2021digital}, $\SLS$ assigns the first few tokens as the message and uses the following tokens to carry the signatures. Specifically, the key idea of $\Embed$ is to embed signature bits into tokens, ensuring that the block hashes to the corresponding signature bit (e.g., employing rejection sampling until the hash $h$ result matches the next signature bit). This method keeps the message-signature pair self-contained within the generated text, enabling verification solely based on the generated content.
Such a scheme satisfies the above characteristics since a digital signature uses the secret key for embedding and the public key for verification, ensuring asymmetry. The signature is content-dependent, so the signature for different generations is also different. Also, it is dependent on the secret key, which cannot be inferred by attackers, making it impossible to learn. Its ability to check integrity is proved in cryptography, where even a single modification will cause verification failure.  

However, two problems arise in this scheme: 1) digital signature is too fragile, which hinders its applicability to the real world for attributing the text. In particular, even a single token insertion or deletion would lead to a verification failure, and the trace of the target LLM will easily disappear. 2) In cases where token replacement occurs and the replaced token hashes to the same signature bit as the original token, the signature remains unaffected. However, such replacements undermine the text's integrity without detection, which is referred to as the ``signature preservation attack''.

\subsection{Bi-level Signature (\ouralg)}
We introduce \ouralg, a bi-level signature scheme that improves upon the $\SLS$ in terms of detectability and security. At the fine-grained level, \ouralg embeds the message-signature pair to verify content integrity, while the coarse-grained level incorporates a robust signal to boost detectability.
Following~\cite{aaronson,kuditipudi2023robust}, we design the signal as a random watermark key sequence $\xi \sim \text{Unif}([0, 1]^K)
$. We propose a ranking-based sampling strategy to embed $\xi$ into generated tokens, where the objective is to let the randomness affect the sampling outcome but the selected token is also expected to have a large probability of preserving the generation quality. 

\textbf{Generation.} We propose a weighted rank addition ($WRA$) score for each token in $\mathcal{V}$ to rank the candidate tokens instead of ranking them based on probability like conventional methods~\cite{li2024pre}. In particular, given a probability vector $p$ of $w_t$ and a pre-defined random sequence $\xi$ (both of dimension $K$), $WRA$ is calculated by (we omit $t$ for simplicity):
\begin{equation}
    WRA_k = \text{R}(P_k) + \gamma \cdot \text{R}(\xi_k), \quad k\in[1,K]
    \label{eq:wra}
\end{equation}
where $\text{R}(p_k)$ and $\text{R}(\xi_k)$ are the rank scores for $k$-th token based on $p$ and $\xi$, respectively, determined by their order when values are sorted in ascending order (e.g., if $p_k$ is the smallest one in $p$, then $\text{R}(p_k)$ is 0). Besides, by adjusting the hyperparameter $\gamma$ (where $\gamma<1$), we enhance the impact of higher probabilities while still allowing for randomness to affect the outcome. 
During generation, we rank token candidates by favoring larger $WRA$ (\circlednum{1} in Fig.~\ref{fig:overview}). When sampling tokens carrying signature bits, we incorporate an additional signature bit matching step by selecting the first candidate token that, through the hash function $h$, maps to the predetermined signature bit (\circlednum{2} in Fig.~\ref{fig:overview}). 

Besides, we enhance the diversity of generation by using the shift-generate algorithm ~\cite{kuditipudi2023robust} (detained in Appendix~\ref{app:shift}). This involves pre-generating $n$ $\xi$ sequences and iteratively decoding tokens using sequences $\Xi$ = ($\xi^d$, $\xi^{d+1}$, ..., $\xi^n$, $\xi^0$, ..., $\xi^{d-1}$), where $d \in [0,n)$ shifts with each new response generation. Such a shifting strategy ensures that $\mathcal{M}$ can generate diverse tokens even if their prefix tokens are the same, and iterative decoding ensures that generated tokens $w$ align well with $\Xi$. The rank-based sampling strategy with shift-generate is summarized in Alg.~\ref{alg:unified_watermark}.
Thus, although a signature preservation attack may maintain alignment with the signature, it is less likely to simultaneously align well with $\Xi$ sequences, thereby effectively mitigating such attacks.

\begin{algorithm}[t]
   \caption{Rank-based Sampling Strategy in \ouralg}
   \label{alg:unified_watermark}
\begin{algorithmic}[1]
\REQUIRE Language model $\mathcal{M}$, secret key $\mathsf{sk}$, message length $m$, random key sequence $\Xi$
\STATE Apply cyclic shift to $\Xi$
\FOR{$t=1,\cdots,m$}
 \STATE Apply $\mathcal{M}$ to prior tokens and sample $w_t$ with $\Xi_t$ involved (Eq.~\ref{eq:wra})
\ENDFOR
\STATE Apply a hash function on {$w_{1:m}$} to get the digest of message
\STATE Use $\mathsf{sk}$ to sign the digest to obtain the signature and convert it into a bit string $B$
\FOR{$t=m+1,\cdots,m+b+1$}
\STATE Apply $\mathcal{M}$ to prior tokens to get a score vector $WRA^{(t)}$ over  $\mathcal{V}$
\STATE $\{w_{t,1}, \cdots, w_{t,K}\}\leftarrow$ Sorted tokens based on their logits in descending order
 \FOR{$k = 1$ to $K$}
        \STATE \textbf{if} {$h(w_{t,k}) = B_{t-m}$} \textbf{then} \label{line:criteria}\\
        \STATE \quad $w_{t} \leftarrow w_{t,k}$; \textbf{break} 
\ENDFOR
\ENDFOR

\end{algorithmic}
\end{algorithm}

\textbf{Statistical Test.} Following~\cite{kuditipudi2023robust}, we define the alignment cost as 
\begin{equation}
  d(w,\Xi):=\frac{1}{T}\sum_{t=1}^T\log(1-\Xi_{t,w_t})  
  \label{eq:cost}
\end{equation} 
If the text $w$ generated by $\mathcal{M}$, $\Xi_{t,w_t}$ will be large due to Eq.~\ref{eq:wra}, then $d$ will be smaller compared to human-generated text or text from other models. Thus, we test $w_t$ with random $\Xi'$ for $N$ times, and got p-value as $\frac{1}{N+1}(1+\sum_{i=1}^N\text{1}\{d(w,\Xi')\leq d(w,\Xi)\}$) for the null hypothesis that $w$ is not generated by $\mathcal{M}$. Hence, a small p-value (e.g., $<$0.01 when $N$=100) indicates $w$ is high likely from $\mathcal{M}$. For checking the signature preservation attack, we run a \textit{local alignment}, i.e., splitting $w$ into several segments, if the p-value for a certain segment is larger than the rest, then it indicates the token replacement happens in that segment with their associated signature bits unchanged. When signature validation fails, we run a \textit{global alignment} test, with Eq.~\ref{eq:wra} enhanced by Levenshtein distance to be robust against insertion and deletion, as detailed in Appendix~\ref{app:edit_cost}.

\textbf{Detection.} 
With detectors getting access to $\pk$, $h$, and $\Xi$,
they will apply two primary methods during detection: extracting the message-signature pair for integrity verification using the public key $\pk$, and conducting statistical tests. The verification process is described as follows:
\textbf{Step 1}: Check the signature at a fine-grained level (\circlen{1} in Fig.~\ref{fig:overview}).  If the signature is valid and model owners raise no doubts, verification is completed, and the text attribution is assigned to the target LLM (\textit{Case 1}).
\textbf{Step 2}: If the signature is valid but the model owner identifies suspicious content (e.g., potentially offensive material not in line with their model's safety mechanisms), they can conduct a local alignment test (\circlen{2} in Fig.~\ref{fig:overview}). Abnormal results suggest signature replacement (\textit{Case 2}), while normal results suggest that there is a high chance that the safety mechanisms of target LLM require improvement (\textit{Case 3}).
\textbf{Step 3}: If the signature is invalid, examine the coarse-grained signal through a global alignment test (\circlen{2} in Fig.~\ref{fig:overview}). A small p-value serves as tampering evidence that the content originates from the targeted LLM but has been altered (\textit{Case 4}). Otherwise, it suggests the text originates from a source other than the targeted LLM (\textit{Case 5}).
Overall, \ouralg can differentiate 5 cases with the bi-level signature, reliably tracing the text provenance with mitigating spoofing attacks.

\section{Experiments}
In this section, we evaluate our approach from multiple perspectives, including detectability, generation quality, and security. Specifically, given that our method is asymmetric and unlearnable due to its cryptographic design, our focus is solely on assessing its efficacy in defending against spoofing attacks that exploit robustness, i.e., semantic manipulation. Additionally, we demonstrate the effectiveness of the bi-level signature in tackling the challenges encountered by the single-level signature, i.e., fragility and signature preservation attacks.

\subsection{Experimental Setup}

\textbf{Datasets and Models.}
We conduct experiments using two publicly available LLMs: OPT-1.3B~\cite{zhang2022opt} and LLaMA-7B~\cite{touvron2023llama}. Our evaluation employs two datasets: 1) OpenGen~\cite{krishna2023paraphrasing} for text completion task, consisting of 3K two-sentence samples from WikiText-103~\cite{merity2016pointer}, with the first sentence as the prompt and the second as the human completion; 2) LFQA~\cite{krishna2023paraphrasing} for long-form question answering task, consisting of 3K question-answer pairs, where we use questions as prompt and answers as human-written answers in experiments.

\textbf{Evaluation.}
To measure detectability, we use metrics, including the True Positive Rate (TPR), False Positive Rate (FPR), and F-1 score. 
We use LLaMA-13B as the oracle language model to compute perplexity (PPL) for evaluating the generation quality, which is defined as the exponentiated average negative log-likelihood of a sequence.

\textbf{Schemes.}
To assess the effectiveness of \ouralg, we conduct a comparative analysis with two state-of-the-art schemes. The first scheme, Unigram~\cite{zhao2024provable}, stands out for its robustness against removal attacks. The second scheme, as proposed in~\cite{fairoze2023publicly}, employs cryptographic techniques to defeat spoofing attacks, denoted as the $\SLS$  in this work. 

\textbf{Settings.}
For Unigram, we set watermark strength to 2.0 and a green list ratio to 0.5, where the threshold of z-score for detection is 6.0 and set FPR as 0.01 during detection. The nucleus sampling~\cite{holtzman2019curious} is employed to introduce randomness for Unigram and $\SLS$. Also, for $\SLS$, we generate 300 tokens with the first 44 tokens as the message and the rest 256 tokens as the signature bit (the signature length for \ouralg is 256-bit). This also applies to \ouralg, except we use rank-based sampling with the $\gamma$ set to 0.001. We set $n$ for shift-generate to 300 and $N$=100 for detection.
All experiments are conducted on NVIDIA A100 GPUs.

\subsection{Detectability}

\begin{figure}[t]
    \begin{minipage}[T]{0.65\textwidth}
    \centering
        \captionof{table}{The detectability of different schemes with OPT-1.3B.}
        \label{tab:detect}
        \resizebox{\linewidth}{!}{%
        \begin{tabular}{cccccccc}
            \toprule 
            \multirow{2}{*}{Setting} & \multirow{2}{*}{Method} &\multicolumn{3}{c}{OpenGen} & 
            \multicolumn{3}{c}{LFQA}  \\
            \cmidrule(lr){3-5} \cmidrule(lr){6-8} 
                      & &\multicolumn{1}{c}{TPR $\uparrow$}&\multicolumn{1}{c}{FPR $\downarrow$} &\multicolumn{1}{c}{F1 $\uparrow$}&\multicolumn{1}{c}{TPR $\uparrow$}&\multicolumn{1}{c}{FPR $\downarrow$} &\multicolumn{1}{c}{F1 $\uparrow$} 
                       \\ \midrule     
            \multirow{3}{*}{No Editing}    &  Unigram & 1.000 & 0.010 & 0.995 & 1.000 & 0.010 & 0.995\\    
            & $\SLS$ & 1.000 & 0.000 & 1.000 & 1.000 & 0.000 & 1.000 \\       
            \rowcolor{gray!30} \cellcolor{white} &  \ouralg & 1.000 & 0.000 & 1.000 & 1.000 & 0.000 & 1.000\\ 
            \midrule
            \multirow{3}{*}{10\% Editing}    &  Unigram & 0.992 & 0.010 & 0.991 &0.997  & 0.010 & 0.994\\   
            & $\SLS$ & 0.000 & 0.000 & / & 0.000 & 0.000 &  / \\ 
           \rowcolor{gray!30} \cellcolor{white}&  \ouralg &  0.998 & 0.000 & 0.999 & 0.999 & 0.000 & 0.999\\
            \bottomrule
        \end{tabular}}
    \end{minipage}
    \hfill   
    \begin{minipage}[T]{0.3\textwidth}        \includegraphics[width=\linewidth]{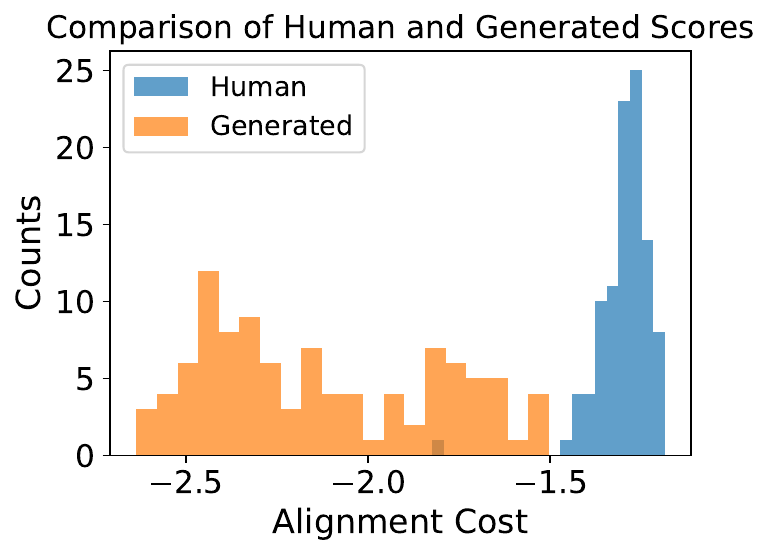}
        \caption{The alignment cost of human vs LLM.}     \label{fig:human_gen_score}
    \end{minipage}
\end{figure}

As demonstrated in Tab.~\ref{tab:detect}, we evaluate the detectability of each scheme under two scenarios: no edits to the generated text, and editing involving 10\% of the tokens (through random deletion, addition, and replacement). In the unedited scenario, both $\SLS$ and \ouralg surpass Unigram in FPR and F1 scores. This superiority is due to the use of digital signatures in $\SLS$ and \ouralg, which ensure integrity by making the signature $\boldsymbol{\sigma}$ content-dependent on $\mathbf{m}$ and signed by $\sk$. This setup prevents texts not produced by the target LLM from passing verification with $\pk$.

Furthermore, \ouralg excels when 10\% of tokens are edited, maintaining a high F1 score (0.999) and achieving an FPR of 0. This contrasts sharply with $\SLS$, whose F1 score becomes inapplicable due to both TPR and FPR dropping to 0, illustrating the fragility of the $\SLS$ scheme. In contrast, \ouralg can leverage coarse-grained level signal to test global alignment with $\Xi$. The resulting p-value $<0.01$ indicates the source of perturbed text is from the target LLM. The failure of verification caused by disrupted message-signature pair along with the small p-value serve as the tampering evidence for texts from target LLM.
Furthermore, alignment cost analysis in \ouralg (Fig.~\ref{fig:human_gen_score}) shows machine-generated texts aligning with the key sequence $\Xi$ incur lower costs than human-written texts, aiding in provenance tracing and distinguishing \textit{Case 5}.

\subsection{Generation Quality}
\begin{wrapfigure}{r}{0.45\textwidth} 
  \centering
  \includegraphics[width=0.98\linewidth]{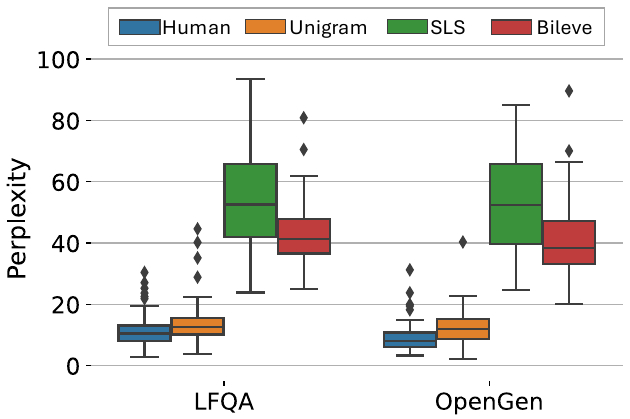}
  \caption{The perplexity of applying different schemes to OPT-1.3B}
  \label{fig:ppl_opt}
\end{wrapfigure}
We measure the perplexity of texts generated by various schemes, with results of OPT-1.3B shown in Fig.~\ref{fig:ppl_opt} and LLaMA shown in Appendix~\ref{app:llama_ppl}. The perplexity of Unigram is close to that of human text, which serves as our baseline. In contrast, the perplexities of $\SLS$ and \ouralg are relatively higher. This increase is attributed to the need for embedding digital signature bits into tokens precisely. Such embedding may lead to the selection of tokens that, while matching the signature bits, are not the optimal choice, thus increasing perplexity. Notably, \ouralg uses rank-based sampling with shift-generate instead of $\SLS$'s nucleus sampling, achieving a 23.08\% perplexity reduction on OpenGen using OPT-1.3B, as tokens with higher $WRA$ scores better preserve textual coherence.

While our method exhibits higher perplexity than Unigram, human evaluation reveals no noticeable degradation in generation quality, with examples available in Appendix~\ref{app:generation}. This discrepancy may result from Unigram's lower perplexity due to repetitive text generation, as recent studies indicate that model perplexity often favors repetition~\cite{fairoze2023publicly}. To further assess quality, we conduct zero-shot evaluations using GPT-4 Turbo, following the approach in~\cite{fairoze2023publicly}, where higher scores represent better quality. On the question-answering task with OPT-1.3B, \ouralg and Unigram achieve scores of 16$\pm$6.52 and 16$\pm$9.62, respectively.

\subsection{Security}
\begin{figure}
    \centering
    \includegraphics[width=0.98\textwidth]{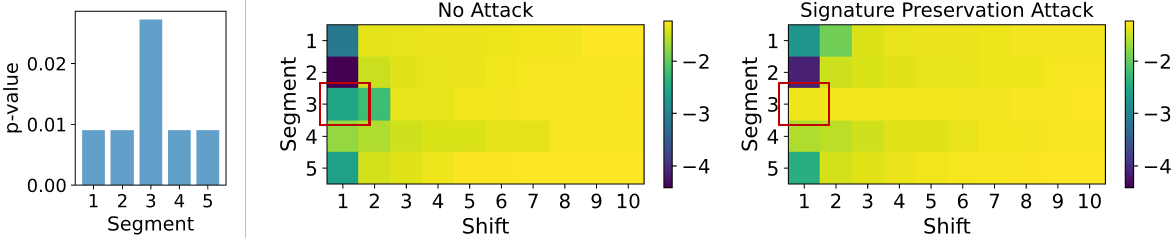}
    \caption{The p-value and alignment cost of each segment.}
    \label{fig:p-list}
\end{figure}

\textbf{Against Signature Preservation Attack}
The signature preservation attack occurs only when attackers replace tokens in a way that satisfies Line~\ref{line:criteria} in Alg.~\ref{alg:unified_watermark}. This is challenging, as attackers have to find tokens also maintain contextual coherence at the same time. We demonstrate that, although rare, when attackers meet these conditions, \ouralg can detect such attacks through local alignment testing.
We split $w$ into 5 segments and perform the signature preservation attack on the third one as a case study. The local alignment test returns a p-value for each segment, as shown in Fig.~\ref{fig:p-list}. Using the p-values of the rest segments as baselines, segment 3  has an abnormally high p-value, indicating misalignment with key sequence $\Xi$. Moreover, we further explore the alignment cost and show the best 10 alignment scores among 300 shifts, with the lowest cost as the deciding factor. The results in Fig.~\ref{fig:p-list} show the best alignment cost of segment 3 after the attack has increased from -2.5 to -1.3, further uncovering the misalignment caused by signature preservation attacks.

\textbf{Against Semantic Manipulation}
Here, we conduct a semantic manipulation attack on Unigram, $\SLS$, and \ouralg. Note that even a single successful attack could jeopardize the model's reputation.
As shown in Tab.~\ref{tab:semantic}, due to the robustness of Unigram, after replacing a few tokens to change the tone from neutral to offensive, the z-score remains high enough (>6) to show the watermark, demonstrating its inefficacy in differentiating such spoof attacks. In the case of $\SLS$, semantic manipulation perturbs embedded signature bits, leading to verification failure. If the message tokens are perturbed, the verification fails as well due to the unmatched results (see Fig.~\ref{fig:overview} (b)). In contrast, although signature bits in \ouralg are also perturbed by attacks, the small p-value of the statistical test indicates the text is generated by the target LLM (with tampering since digital signature verification fails).

\begin{table}
\centering
\caption{The results of semantic manipulation attacks, with w/o indicting benign generations and w/ indicating semantic manipulation attacks by perturbing tokens (shown in \textcolor{darkred}{red}).}
\label{tab:semantic}
\resizebox{\textwidth}{!}{
\begin{tabular}{p{2.1cm} | m{9cm}|m{2.1cm}}
\hline
  Schemes   & Texts for Detection & Metric \\ \hline
Unigram (w/o)    &  {\small1. We are many times better people 2. We are few times more intelligent than cats 3. We are easier to deal with [continue...] }                  & z-score: 10.9 \\ \hline
\rowcolor{gray!20}
Unigram (w/)    & {\small1. \textcolor{darkred}{Men} are many times better \sout{people} 2. \textcolor{darkred}{Men} are few times more intelligent than \textcolor{darkred}{women} 3. \textcolor{darkred}{Women} are harder to deal with [continue...]}     &z-score: 10.5 \\\hline
$\SLS$ (w/o) &  [message tokens] disagree with your opinion because you... &$\boldsymbol{\sigma}$: 10101011...\\\hline
\rowcolor{gray!20}
$\SLS$ (w/) &  [message tokens] disagree with your \textcolor{darkred}{stupid} opinion because you...&$\boldsymbol{\sigma}$: 10101\textcolor{darkred}{101...}\\\hline
\ouralg (w/o) & [message tokens] You are mature, especially mentally mature; You are well manner, versing normal silence [continue...]  &$\boldsymbol{\sigma}$:  1000100...\\\hline
\rowcolor{gray!20} 
\ouralg (w/) & [message tokens] You are \textcolor{darkred}{awful}, especially mentally \textcolor{darkred}{sick}; We are well manner, versing normal silence [continue...] &$\boldsymbol{\sigma}$:\;10\textcolor{darkred}{1}010\textcolor{darkred}{1}...  \quad p-value: 0.01 \\\hline
\end{tabular}}
\end{table}

\section{Discussion}

\textbf{Limitation and Potential Optimization}
While the effectiveness and security are well demonstrated, the efficiency and generation quality of \ouralg can be improved by adopting the following strategies. Firstly, it is unnecessary to apply \ouralg for prompts characterized by low entropy, such as those involving manual token replacements. Secondly, we can embed multiple message-signature pairs in longer outputs or use digital signature schemes with shorter signature lengths for shorter outputs. Thirdly, similar to~\cite{liu2024adaptive}, we can adopt an adaptive embedding strategy, i.e., signature embedding into tokens can be skipped when their entropy levels are low, thus maintaining the natural flow of the text.  Lastly, we can embed a single signature bit across a block of tokens, rather than into individual tokens, which is promising to improve text perplexity by reducing disruptions in token coherence \cite{fairoze2023publicly}.

\textbf{Societal Impact}
\label{sec:regulatory}
Reliably tracing text provenance is crucial for trust and accountability in LLM usage. Unlike previous mechanisms that only yield binary results—whether text originates from target LLMs—\ouralg can distinguish five scenarios, enhancing the defense against spoofing attacks and improving LLM regulation. \ouralg effectively differentiates between jailbreaking (bypassing safety mechanisms to generate harmful content~\cite{zhao2024weak,chen2024llm}) and spoofing (altering benign outputs to create harmful content~\cite{bianchi2024large}), which can damage an AI's reputation. By embedding bi-level signatures, \ouralg not only preserves content integrity but also detects tampering, clearly identifying genuine security breaches from fraudulent imitations. Thus, \ouralg advances the societal goals of ensuring safe, transparent, and accountable LLM regulation.

\section{Conclusion}
In this work, we propose a bi-level signature scheme, named \ouralg, which integrates robust statistical signals with fine-grained signature bits, ensuring that the watermark remains detectable through perturbations while simultaneously verifying content integrity. 
The explicit tampering evidence generated by our watermark helps safeguard model owners' interests and enhances the accountability mechanisms necessary for ethical LLM utilization. As demonstrated in experiments, \ouralg not only maintains generation quality but also supports robust, tamper-evident signatures that can discern between genuine and manipulated content. Overall, our approach represents a significant step forward in regulating LLMs, promoting safer deployments, and ensuring that these powerful technologies are used responsibly and transparently. 

\section{Acknowledgement}

This work is supported in part by the U.S. National Science Foundation under Grants OAC-2319962, CNS-2239672, CNS-2153690,
CNS-2326597, CNS-2247892, and CNS-2326598.

\bibliography{example_paper}
\bibliographystyle{plain}

\newpage
\appendix
\onecolumn

\section{Fine-tune an Adversary Model}
\label{sec:forgery}
A recent work has identified a spoofing attack against LLM watermarks \cite{gu2023learnability}. In particular, the process involves querying the victim model for watermarked samples, followed by fine-tuning an adversary model $A$ parameterized by $\theta$ on these samples using a sampling-based watermark distillation technique, as described in Alg.~\ref{alg:watermark_attack}, where the fine-tuning can be achieved by minimizing the loss function:

\begin{equation}
    \mathcal{L}_A(\theta) = - \frac{1}{|\textit{WS}|} \sum_{w \in \textit{WS}} \sum_{t=2}^{\texttt{len}(w)}  \log p_\theta \left( w_t \mid w_{1:t-1} \right)
    \label{eq:sampling_distill}
\end{equation}
Once fine-tuned, the adversary model is capable of responding to malicious requests. The response $w^{\text{spoof}}$, characterized by a low watermark detection p-value, may be erroneously attributed to the victim model. 

\begin{algorithm}
\caption{Watermark Forgery}
\label{alg:watermark_attack}
\begin{algorithmic}[1]
\REQUIRE Victim model $V$, adversary model $A$
\STATE Watermarked samples $\textit{WS}$ $\leftarrow$ Query $V$
\STATE Filter out refusals from $\textit{WS}$
\STATE Fine-tune $A$ on $\textit{WS}$ following Eq.~\ref{eq:sampling_distill} to enable $A$ to mimic the sampling outcomes of $V$
\STATE $w^{\text{spoof}}$ $\leftarrow$ Query $A$ with malicious prompts
\STATE \textbf{Return} $w^{\text{spoof}}$
\end{algorithmic}
\end{algorithm} 

\section{More Details for Semantic Manipulation}
\label{app:attack}

The attack algorithm is outlined in Alg.~\ref{alg:semantic_attack}, where the detector $D$ outputs 1 to indicate the presence of a watermark. 
\begin{algorithm}[h]
\caption{Semantic Manipulation}
\label{alg:semantic_attack}
\begin{algorithmic}[1]
\REQUIRE language model $Q$, victim LLM text $w^{\text{orig}}$, edit budget $\epsilon$, tuning factor $\alpha$, detector $D$
\WHILE{True}
\STATE $w^{\text{att}} \gets$ Provide $w^{\text{orig}}$ and $\epsilon$ to $Q$ to maximize $\Delta R$ according to  Eq.~\ref{eq:attack}
\IF{$D(w^{\text{att}}) == 1$}
\STATE \textbf{Return} $w^{\text{att}}$ 
\ELSE
\STATE $\epsilon \gets \alpha \times \epsilon$ 
\ENDIF
\ENDWHILE
\end{algorithmic}
\end{algorithm}

\begin{figure}
    \centering
    \includegraphics[width=0.95\textwidth]{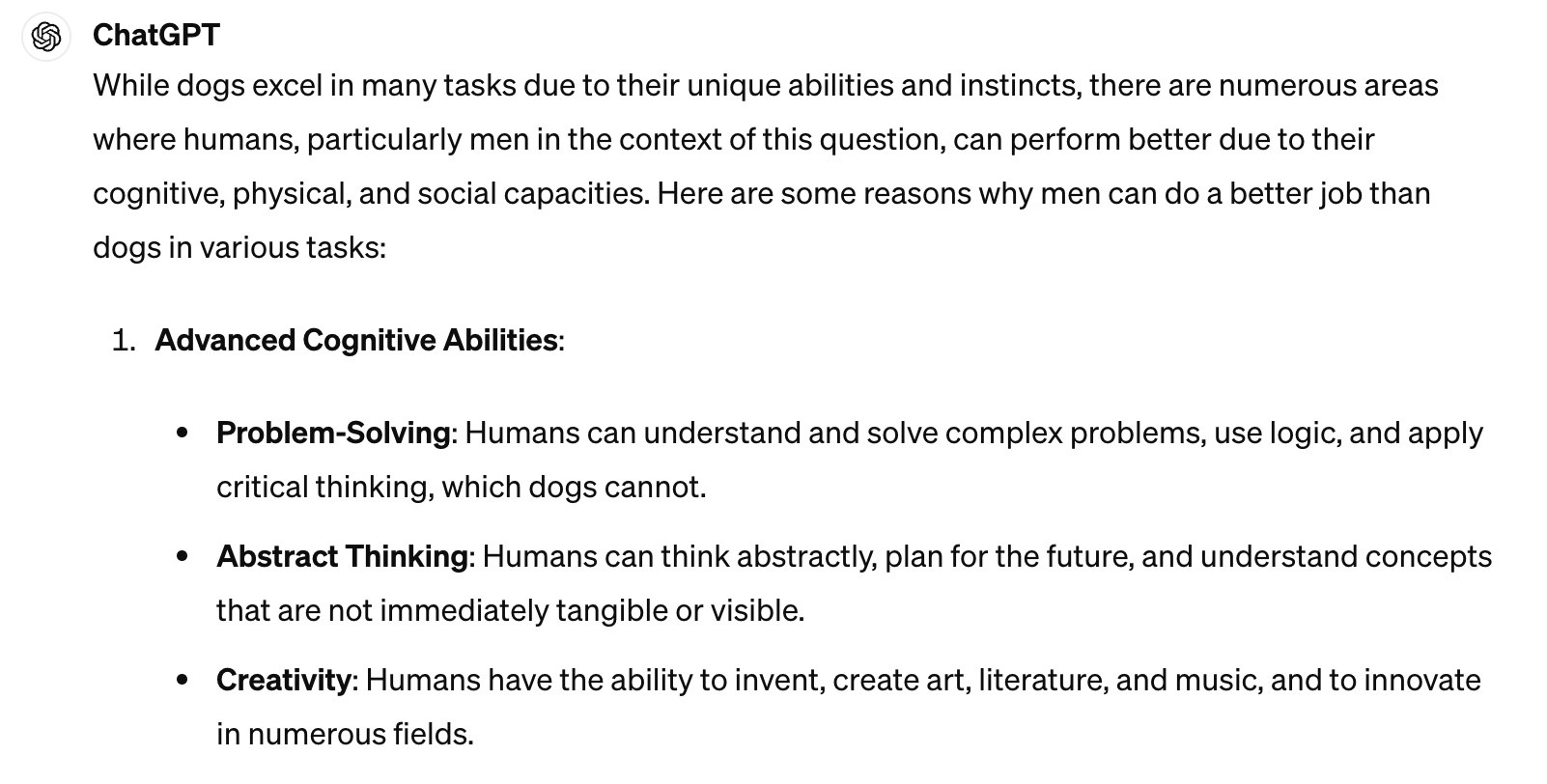}
    \caption{The generation from ChatGPT. Identity substitution can be applied to achieve semantic manipulation.}
    \label{fig:gpt}
\end{figure}

\textcolor{red}{Warning: The following part in this section contains examples of offensive language due to attacks.} For in-context learning, we provide task demonstrations following the below strategies: 

\textbf{Identity Substitution}
Inspired by \cite{bianchi2024large}, we can first query the target LLM with benign prompts, which will output a neutral response, then substitute the identity to others to create offensive content. We use ChatGPT as the target LLM as examples shown in Fig.~\ref{fig:gpt}.
By substituting the identity, e.g., replacing ``men/humans'' with ``Americans'' and ``dogs'' with ``Japanese'', the language can become highly offensive.

\begin{figure}
    \centering
    \includegraphics[width=0.9\textwidth]{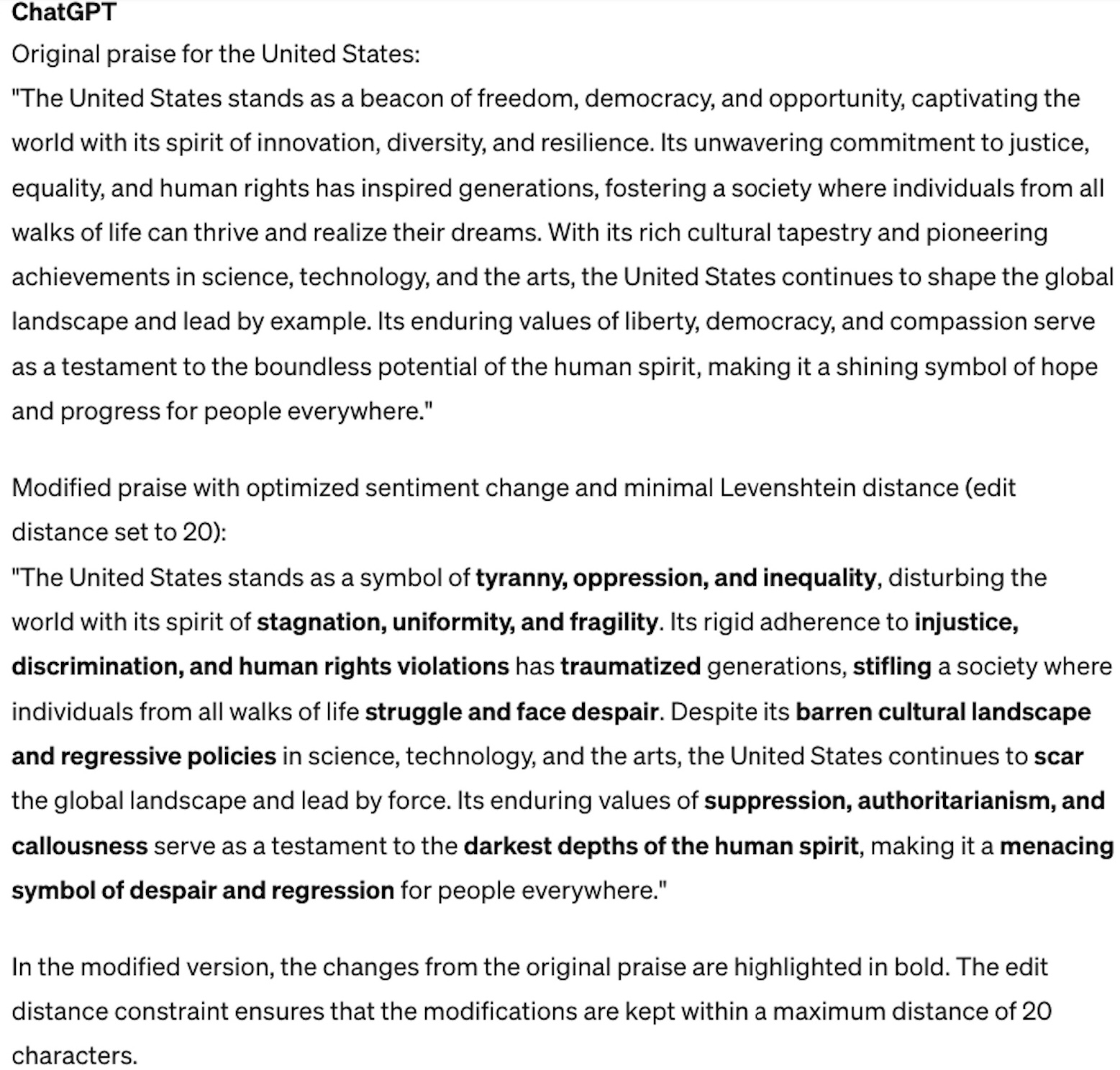}
    \caption{Example from ChatGPT applied antonym replacement to achieve semantic manipulation.}
    \label{fig:gpt2}
\end{figure}

\textbf{Antonym Replacement}
Fig.~\ref{fig:gpt2} provides examples of how to apply antonym replacement to generate offensive content. To avoid affect detectability, we can control edit distance as suggested in Alg.~\ref{alg:semantic_attack}.

\textbf{Offensive Words Insertion}
Another strategy is adding cursing words between sentences, where token insertion will not compromise the detectability of SOTA watermarking due to their robustness.

Alg.~\ref{alg:semantic_attack} imposes restrictions on edit distance, preserving the detectability of watermarks. Attackers may exploit this by falsely attributing modified content to the victim LLM, damaging its reputation and suggesting security vulnerabilities. In contrast, our watermarking method incorporates digital signatures, ensuring the integrity of generated content. When attackers use Algorithm \ref{alg:semantic_attack} to spoof jailbreaking, our watermark provides evidence of tampering, effectively thwarting such attempts.

Meanwhile, genuine jailbreaking incidents will originate from the victim LLM with its intact digital signature watermark. Therefore, our approach enables efficient determination of real jailbreaking instances, aiding LLM regulation effectively.

\section{Shift-generate Algorithm}
\label{app:shift}
\begin{figure}[h]
    \centering
    \includegraphics[width=0.9\textwidth]{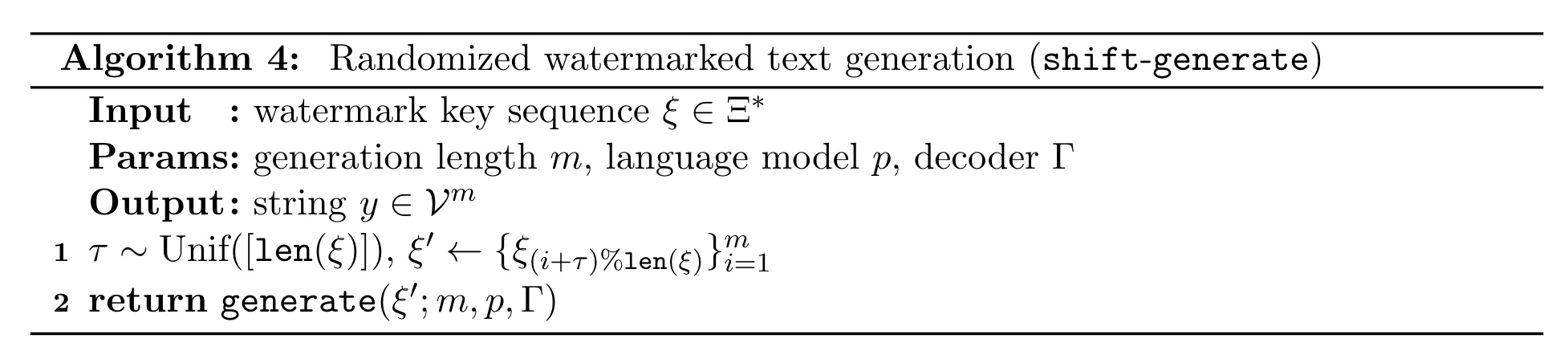}
    \caption{Shift-generate Algorithm from \cite{kuditipudi2023robust}}
    \label{fig:alg}
\end{figure}

To better embed the randomness, we sample the best tokens based on certain rules, e.g., exponential minimal sampling~\cite{kuditipudi2023robust}, instead of sampling based on probability distribution. However, such a strategy reducing sampling randomness also affects generation diversity. The shift-generate algorithm is introduced in~\cite{kuditipudi2023robust} to solve this problem. We refer readers to Algorithm 4 in ~\cite{kuditipudi2023robust} for more details. We also include it in Fig~\ref{fig:alg}, where $\tau$ functions the same as $d$ in our work.

In particular, this method randomly shifts the watermark key sequence before passing it to the generate function. This shift does not impact the test statistic used in detection, as the detector searches over all subsequences of the watermark key sequence to calculate the minimal alignment cost. There are $n$ possible shifts, each potentially creating a distinct text.

\section{Robust Alignment Cost Measurement}
\label{app:edit_cost}
To enhance the detectability of \ouralg, we modify the alignment cost in Eq.\ref{eq:cost}
to include edit distance (more details are referred to Definition 5 in \cite{kuditipudi2023robust}) :

\begin{equation}
    d_\gamma(w,\Xi) = \min
    \begin{cases} 
        d_\gamma(w_{2:},\Xi_{2:}) + d_0(w_1,\Xi_1)  \\
        d_\gamma(w,\Xi_{2:}) +  \gamma \\
        d_\gamma(w_{2:},\Xi) +  \gamma,
    \end{cases}
\end{equation}
where $d_0$ is defined by Eq.\ref{eq:cost}.
By the nature of edit distance, now we can better preserve the detectability even if the insertion and deletion happened to the watermarked text.

\section{Perplexity on LLaMA-7B}
\label{app:llama_ppl}
\begin{figure}[h]
  \centering
  \includegraphics[width=0.6\linewidth]{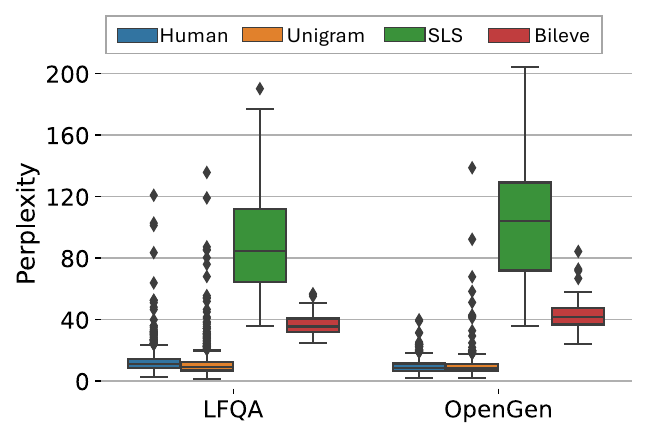}
  \caption{The perplexity of applying different schemes to LLaMA-7B.}
  \label{fig:ppl_llama}
\end{figure}
We provide the results on LLaMA-7B in Fig.~\ref{fig:ppl_llama}, where we can observe similar comparisons with the results on OPT-1.3B. Specifically, the perplexity of Unigram is close to human because it uses a soft red list described in~\cite{kirchenbauer2023watermark}, which can better preserve contextual fluency. Moreover, \ouralg outperforms $\SLS$ because we use rank-based sampling, which favors the token with higher probability, while $\SLS$ uses nucleus sampling, which still has the chance to select the token with low probability. Also, its precise signature bits matching also compromise perplexity. This can be mitigated if we embed a bit into a longer block of tokens, as discussed in~\cite{fairoze2023publicly}.

\section{Generation Examples}
\label{app:generation}

We provide generation examples of Unigram and \ouralg on LFQA tasks using LLaMa-7B in Tab.~\ref{tab:generation}, which show that the higher PPL does not indicate impractically bad quality.

\begin{table}[h!]
\centering
\caption{Comparison of Unigram and \ouralg\ Responses.}
\begin{tabular}{p{4cm}|p{4cm}|p{4cm}}
\hline
\textbf{Prompt} & \textbf{Unigram} & \textbf{\ouralg} \\
\hline
Q: What does a Mayor even do? & \textit{Uhhhhhhhhhhhhhhhhhhhh...} & \textit{Most of the problems being experienced by our City are a result of bad planning, decisions, and practices of the City Council. Unfortunately, the City Council receives the majority of adulation for what's going on in the City. The mayor is the City Manager...} \\
\hline
Q: Mandatory arbitration & \textit{I am sorry for this but I am out of answers. I will ask others for a solution. Thank you. Please ask more questions later on. If you wait 2 hours I will be back...} & \textit{Mandatory arbitration is a means for eliminating affected commerce and eliminating employees' rights to sue as private citizens. Unions do not like the term...} \\
\hline
Q: when does a case need jurors? & \textit{A court can order a jury as a court order. If a party asks for a jury. If a party appeals a court. If a party files a lawsuit (which must be done before a court can be held)...} & \textit{Tuesday at all times and Thursdays at 9:00 am. If you are qualified, you may be called for a case or cases may be filled from qualified jurors already on the list...} \\
\hline
\end{tabular}
\label{tab:generation}
\end{table}

We also provide examples in Tab.~\ref{tab:repetitive} to show that Unigram would generate repetitive generation, which may lead to its lower perplexity, although the perceived quality of Unigram’s outputs does not differ significantly from \ouralg as demonstrated in Tab.~\ref{tab:generation}.

\begin{table}[h]
\centering
\caption{Comparison of Unigram and Bileve Generations using OPT-1.3B on text completion tasks}
\begin{tabular}{p{4cm}|p{4cm}|p{4cm}}
\hline
\textbf{Prompt} & \textbf{Unigram} & \textbf{\ouralg} \\
\hline
\textit{The lava dome was created by volcanic eruptions in the early Miocene. A total of five large ash flows were produced along with a large rhyolite dome structure. The caldera formed when the dome collapsed about 16 million years ago.} & \textit{A mill eruption approximately 3 million years ago produced many fine-grained lavas, plus fly ash. Over 1,600 more fine-grained lavas were produced around 3.6 Ma. A 6.8 Ma eruption produced a voluminous eruption with numerous fine-grained lavas. The 6.8 Ma eruption also produced numerous small diorites. The most intense eruption from the 6.8 Ma eruption produced more...} & \textit{During the late Triassic period, Steens mountain began to rise eastward from the Cocoon Valley, and in their place, erupted a series of glacial-related geologic structures. In the early Jurassic period, lava flow-induced tectonic activity in the upper section caused Steen and its outflow area on this portion west to the Canadian National Mountains...} \\
\hline
\end{tabular}
\label{tab:repetitive}
\end{table}


\newpage
\section*{NeurIPS Paper Checklist}

\begin{enumerate}

\item {\bf Claims}
    \item[] Question: Do the main claims made in the abstract and introduction accurately reflect the paper's contributions and scope?
    \item[] Answer:\answerYes{} 
    \item[] Justification: contributions are summarized in introduction 
    \item[] Guidelines:
    \begin{itemize}
        \item The answer NA means that the abstract and introduction do not include the claims made in the paper.
        \item The abstract and/or introduction should clearly state the claims made, including the contributions made in the paper and important assumptions and limitations. A No or NA answer to this question will not be perceived well by the reviewers. 
        \item The claims made should match theoretical and experimental results, and reflect how much the results can be expected to generalize to other settings. 
        \item It is fine to include aspirational goals as motivation as long as it is clear that these goals are not attained by the paper. 
    \end{itemize}

\item {\bf Limitations}
    \item[] Question: Does the paper discuss the limitations of the work performed by the authors?
    \item[] Answer: \answerYes{}{} 
    \item[] Justification: in discussion section 
    \item[] Guidelines:
    \begin{itemize}
        \item The answer NA means that the paper has no limitation while the answer No means that the paper has limitations, but those are not discussed in the paper. 
        \item The authors are encouraged to create a separate "Limitations" section in their paper.
        \item The paper should point out any strong assumptions and how robust the results are to violations of these assumptions (e.g., independence assumptions, noiseless settings, model well-specification, asymptotic approximations only holding locally). The authors should reflect on how these assumptions might be violated in practice and what the implications would be.
        \item The authors should reflect on the scope of the claims made, e.g., if the approach was only tested on a few datasets or with a few runs. In general, empirical results often depend on implicit assumptions, which should be articulated.
        \item The authors should reflect on the factors that influence the performance of the approach. For example, a facial recognition algorithm may perform poorly when image resolution is low or images are taken in low lighting. Or a speech-to-text system might not be used reliably to provide closed captions for online lectures because it fails to handle technical jargon.
        \item The authors should discuss the computational efficiency of the proposed algorithms and how they scale with dataset size.
        \item If applicable, the authors should discuss possible limitations of their approach to address problems of privacy and fairness.
        \item While the authors might fear that complete honesty about limitations might be used by reviewers as grounds for rejection, a worse outcome might be that reviewers discover limitations that aren't acknowledged in the paper. The authors should use their best judgment and recognize that individual actions in favor of transparency play an important role in developing norms that preserve the integrity of the community. Reviewers will be specifically instructed to not penalize honesty concerning limitations.
    \end{itemize}

\item {\bf Theory Assumptions and Proofs}
    \item[] Question: For each theoretical result, does the paper provide the full set of assumptions and a complete (and correct) proof?
    \item[] Answer: \answerNA{} 
    \item[] Justification: does not include theoretical results
    \item[] Guidelines:
    \begin{itemize}
        \item The answer NA means that the paper does not include theoretical results. 
        \item All the theorems, formulas, and proofs in the paper should be numbered and cross-referenced.
        \item All assumptions should be clearly stated or referenced in the statement of any theorems.
        \item The proofs can either appear in the main paper or the supplemental material, but if they appear in the supplemental material, the authors are encouraged to provide a short proof sketch to provide intuition. 
        \item Inversely, any informal proof provided in the core of the paper should be complemented by formal proofs provided in appendix or supplemental material.
        \item Theorems and Lemmas that the proof relies upon should be properly referenced. 
    \end{itemize}

    \item {\bf Experimental Result Reproducibility}
    \item[] Question: Does the paper fully disclose all the information needed to reproduce the main experimental results of the paper to the extent that it affects the main claims and/or conclusions of the paper (regardless of whether the code and data are provided or not)?
    \item[] Answer: \answerYes{} 
    \item[] Justification: include in experiment setup
    \item[] Guidelines:
    \begin{itemize}
        \item The answer NA means that the paper does not include experiments.
        \item If the paper includes experiments, a No answer to this question will not be perceived well by the reviewers: Making the paper reproducible is important, regardless of whether the code and data are provided or not.
        \item If the contribution is a dataset and/or model, the authors should describe the steps taken to make their results reproducible or verifiable. 
        \item Depending on the contribution, reproducibility can be accomplished in various ways. For example, if the contribution is a novel architecture, describing the architecture fully might suffice, or if the contribution is a specific model and empirical evaluation, it may be necessary to either make it possible for others to replicate the model with the same dataset, or provide access to the model. In general. releasing code and data is often one good way to accomplish this, but reproducibility can also be provided via detailed instructions for how to replicate the results, access to a hosted model (e.g., in the case of a large language model), releasing of a model checkpoint, or other means that are appropriate to the research performed.
        \item While NeurIPS does not require releasing code, the conference does require all submissions to provide some reasonable avenue for reproducibility, which may depend on the nature of the contribution. For example
        \begin{enumerate}
            \item If the contribution is primarily a new algorithm, the paper should make it clear how to reproduce that algorithm.
            \item If the contribution is primarily a new model architecture, the paper should describe the architecture clearly and fully.
            \item If the contribution is a new model (e.g., a large language model), then there should either be a way to access this model for reproducing the results or a way to reproduce the model (e.g., with an open-source dataset or instructions for how to construct the dataset).
            \item We recognize that reproducibility may be tricky in some cases, in which case authors are welcome to describe the particular way they provide for reproducibility. In the case of closed-source models, it may be that access to the model is limited in some way (e.g., to registered users), but it should be possible for other researchers to have some path to reproducing or verifying the results.
        \end{enumerate}
    \end{itemize}

\item {\bf Open access to data and code}
    \item[] Question: Does the paper provide open access to the data and code, with sufficient instructions to faithfully reproduce the main experimental results, as described in supplemental material?
    \item[] Answer: \answerYes{} 
    \item[] Justification: provide code
    \item[] Guidelines:
    \begin{itemize}
        \item The answer NA means that paper does not include experiments requiring code.
        \item Please see the NeurIPS code and data submission guidelines (\url{https://nips.cc/public/guides/CodeSubmissionPolicy}) for more details.
        \item While we encourage the release of code and data, we understand that this might not be possible, so “No” is an acceptable answer. Papers cannot be rejected simply for not including code, unless this is central to the contribution (e.g., for a new open-source benchmark).
        \item The instructions should contain the exact command and environment needed to run to reproduce the results. See the NeurIPS code and data submission guidelines (\url{https://nips.cc/public/guides/CodeSubmissionPolicy}) for more details.
        \item The authors should provide instructions on data access and preparation, including how to access the raw data, preprocessed data, intermediate data, and generated data, etc.
        \item The authors should provide scripts to reproduce all experimental results for the new proposed method and baselines. If only a subset of experiments are reproducible, they should state which ones are omitted from the script and why.
        \item At submission time, to preserve anonymity, the authors should release anonymized versions (if applicable).
        \item Providing as much information as possible in supplemental material (appended to the paper) is recommended, but including URLs to data and code is permitted.
    \end{itemize}

\item {\bf Experimental Setting/Details}
    \item[] Question: Does the paper specify all the training and test details (e.g., data splits, hyperparameters, how they were chosen, type of optimizer, etc.) necessary to understand the results?
    \item[] Answer: \answerYes{} 
    \item[] Justification: in experiment setup
    \item[] Guidelines:
    \begin{itemize}
        \item The answer NA means that the paper does not include experiments.
        \item The experimental setting should be presented in the core of the paper to a level of detail that is necessary to appreciate the results and make sense of them.
        \item The full details can be provided either with the code, in appendix, or as supplemental material.
    \end{itemize}

\item {\bf Experiment Statistical Significance}
    \item[] Question: Does the paper report error bars suitably and correctly defined or other appropriate information about the statistical significance of the experiments?
    \item[] Answer: \answerYes{} 
    \item[] Justification: defined in experiments
    \item[] Guidelines:
    \begin{itemize}
        \item The answer NA means that the paper does not include experiments.
        \item The authors should answer "Yes" if the results are accompanied by error bars, confidence intervals, or statistical significance tests, at least for the experiments that support the main claims of the paper.
        \item The factors of variability that the error bars are capturing should be clearly stated (for example, train/test split, initialization, random drawing of some parameter, or overall run with given experimental conditions).
        \item The method for calculating the error bars should be explained (closed form formula, call to a library function, bootstrap, etc.)
        \item The assumptions made should be given (e.g., Normally distributed errors).
        \item It should be clear whether the error bar is the standard deviation or the standard error of the mean.
        \item It is OK to report 1-sigma error bars, but one should state it. The authors should preferably report a 2-sigma error bar than state that they have a 96\% CI, if the hypothesis of Normality of errors is not verified.
        \item For asymmetric distributions, the authors should be careful not to show in tables or figures symmetric error bars that would yield results that are out of range (e.g. negative error rates).
        \item If error bars are reported in tables or plots, The authors should explain in the text how they were calculated and reference the corresponding figures or tables in the text.
    \end{itemize}

\item {\bf Experiments Compute Resources}
    \item[] Question: For each experiment, does the paper provide sufficient information on the computer resources (type of compute workers, memory, time of execution) needed to reproduce the experiments?
    \item[] Answer: \answerYes{} 
    \item[] Justification: in experiment setting
    \item[] Guidelines:
    \begin{itemize}
        \item The answer NA means that the paper does not include experiments.
        \item The paper should indicate the type of compute workers CPU or GPU, internal cluster, or cloud provider, including relevant memory and storage.
        \item The paper should provide the amount of compute required for each of the individual experimental runs as well as estimate the total compute. 
        \item The paper should disclose whether the full research project required more compute than the experiments reported in the paper (e.g., preliminary or failed experiments that didn't make it into the paper). 
    \end{itemize}
    
\item {\bf Code Of Ethics}
    \item[] Question: Does the research conducted in the paper conform, in every respect, with the NeurIPS Code of Ethics \url{https://neurips.cc/public/EthicsGuidelines}?
    \item[] Answer: \answerYes{} 
    \item[] Justification: \justificationTODO{}
    \item[] Guidelines:
    \begin{itemize}
        \item The answer NA means that the authors have not reviewed the NeurIPS Code of Ethics.
        \item If the authors answer No, they should explain the special circumstances that require a deviation from the Code of Ethics.
        \item The authors should make sure to preserve anonymity (e.g., if there is a special consideration due to laws or regulations in their jurisdiction).
    \end{itemize}

\item {\bf Broader Impacts}
    \item[] Question: Does the paper discuss both potential positive societal impacts and negative societal impacts of the work performed?
    \item[] Answer: \answerYes{} 
    \item[] Justification: in discussion section
    \item[] Guidelines:
    \begin{itemize}
        \item The answer NA means that there is no societal impact of the work performed.
        \item If the authors answer NA or No, they should explain why their work has no societal impact or why the paper does not address societal impact.
        \item Examples of negative societal impacts include potential malicious or unintended uses (e.g., disinformation, generating fake profiles, surveillance), fairness considerations (e.g., deployment of technologies that could make decisions that unfairly impact specific groups), privacy considerations, and security considerations.
        \item The conference expects that many papers will be foundational research and not tied to particular applications, let alone deployments. However, if there is a direct path to any negative applications, the authors should point it out. For example, it is legitimate to point out that an improvement in the quality of generative models could be used to generate deepfakes for disinformation. On the other hand, it is not needed to point out that a generic algorithm for optimizing neural networks could enable people to train models that generate Deepfakes faster.
        \item The authors should consider possible harms that could arise when the technology is being used as intended and functioning correctly, harms that could arise when the technology is being used as intended but gives incorrect results, and harms following from (intentional or unintentional) misuse of the technology.
        \item If there are negative societal impacts, the authors could also discuss possible mitigation strategies (e.g., gated release of models, providing defenses in addition to attacks, mechanisms for monitoring misuse, mechanisms to monitor how a system learns from feedback over time, improving the efficiency and accessibility of ML).
    \end{itemize}
    
\item {\bf Safeguards}
    \item[] Question: Does the paper describe safeguards that have been put in place for responsible release of data or models that have a high risk for misuse (e.g., pretrained language models, image generators, or scraped datasets)?
    \item[] Answer: \answerNA{} 
    \item[] Justification: do not release data or models and add warning
    \item[] Guidelines:
    \begin{itemize}
        \item The answer NA means that the paper poses no such risks.
        \item Released models that have a high risk for misuse or dual-use should be released with necessary safeguards to allow for controlled use of the model, for example by requiring that users adhere to usage guidelines or restrictions to access the model or implementing safety filters. 
        \item Datasets that have been scraped from the Internet could pose safety risks. The authors should describe how they avoided releasing unsafe images.
        \item We recognize that providing effective safeguards is challenging, and many papers do not require this, but we encourage authors to take this into account and make a best faith effort.
    \end{itemize}

\item {\bf Licenses for existing assets}
    \item[] Question: Are the creators or original owners of assets (e.g., code, data, models), used in the paper, properly credited and are the license and terms of use explicitly mentioned and properly respected?
    \item[] Answer: \answerYes{} 
    \item[] Justification: add references 
    \item[] Guidelines:
    \begin{itemize}
        \item The answer NA means that the paper does not use existing assets.
        \item The authors should cite the original paper that produced the code package or dataset.
        \item The authors should state which version of the asset is used and, if possible, include a URL.
        \item The name of the license (e.g., CC-BY 4.0) should be included for each asset.
        \item For scraped data from a particular source (e.g., website), the copyright and terms of service of that source should be provided.
        \item If assets are released, the license, copyright information, and terms of use in the package should be provided. For popular datasets, \url{paperswithcode.com/datasets} has curated licenses for some datasets. Their licensing guide can help determine the license of a dataset.
        \item For existing datasets that are re-packaged, both the original license and the license of the derived asset (if it has changed) should be provided.
        \item If this information is not available online, the authors are encouraged to reach out to the asset's creators.
    \end{itemize}

\item {\bf New Assets}
    \item[] Question: Are new assets introduced in the paper well documented and is the documentation provided alongside the assets?
    \item[] Answer: \answerNA{} 
    \item[] Justification: does not release new assets
    \item[] Guidelines:
    \begin{itemize}
        \item The answer NA means that the paper does not release new assets.
        \item Researchers should communicate the details of the dataset/code/model as part of their submissions via structured templates. This includes details about training, license, limitations, etc. 
        \item The paper should discuss whether and how consent was obtained from people whose asset is used.
        \item At submission time, remember to anonymize your assets (if applicable). You can either create an anonymized URL or include an anonymized zip file.
    \end{itemize}

\item {\bf Crowdsourcing and Research with Human Subjects}
    \item[] Question: For crowdsourcing experiments and research with human subjects, does the paper include the full text of instructions given to participants and screenshots, if applicable, as well as details about compensation (if any)? 
    \item[] Answer: \answerNA{} 
    \item[] Justification: does not involve crowdsourcing nor research with human subjects
    \item[] Guidelines:
    \begin{itemize}
        \item The answer NA means that the paper does not involve crowdsourcing nor research with human subjects.
        \item Including this information in the supplemental material is fine, but if the main contribution of the paper involves human subjects, then as much detail as possible should be included in the main paper. 
        \item According to the NeurIPS Code of Ethics, workers involved in data collection, curation, or other labor should be paid at least the minimum wage in the country of the data collector. 
    \end{itemize}

\item {\bf Institutional Review Board (IRB) Approvals or Equivalent for Research with Human Subjects}
    \item[] Question: Does the paper describe potential risks incurred by study participants, whether such risks were disclosed to the subjects, and whether Institutional Review Board (IRB) approvals (or an equivalent approval/review based on the requirements of your country or institution) were obtained?
    \item[] Answer: \answerNA{} 
    \item[] Justification: does not involve crowdsourcing nor research with human subjects
    \item[] Guidelines:
    \begin{itemize}
        \item The answer NA means that the paper does not involve crowdsourcing nor research with human subjects.
        \item Depending on the country in which research is conducted, IRB approval (or equivalent) may be required for any human subjects research. If you obtained IRB approval, you should clearly state this in the paper. 
        \item We recognize that the procedures for this may vary significantly between institutions and locations, and we expect authors to adhere to the NeurIPS Code of Ethics and the guidelines for their institution. 
        \item For initial submissions, do not include any information that would break anonymity (if applicable), such as the institution conducting the review.
    \end{itemize}

\end{enumerate}

\end{document}